# Quantum Message Authentication Based on Classical NP-Complete Problem

Li Yang[*], Lei Hu, and Deng-Guo Feng

State Key Laboratory of Information Security (Graduate School of Chinese Academy of Sciences),
Beijing 100039, P. R. China

This note presents a method to authenticate a quantum message based on the idea of classical SN-S authentication code and the computations between different quantum registers. If the pre-coding generator matrix in SN-S code is public, the quantum scheme is a public-key data integrity scheme; if it is secret, the quantum scheme is a hybrid data origin authentication scheme.

Authentication of quantum message [1] is a basic aspect of quantum cryptology. This note presents a scheme based on a classical NP-Complete problem for authenticating a quantum message.

Consider the original SN-S authentication code[2]. Suppose Generator matrix $G_s$ is $k$ by $n_1$ matrix and in standard form: $G_s = [I_k | A]$, where $I_k$ is the $k$ by $k$ identity matrix, $A$ is chosen randomly form $k$ by $n_1 - k$ matrix. The $[n_1, k]$ linear code generated by $G_s$ need not be of any error-correcting or error-detecting capability. Generalized inverse matrix $G_s^-$ satisfies: $G_s G_s^- = I_k$. Public matrix is $G' = SGP$, where $G$ is a generator matrix of a Goppa code $C$, S is an invertible matrix, and P is a permutation matrix. ($S, G, P$) is the private key of the receiver Bob. The generalized inverse matrix $G'^-$ satisfies: $G'G'^- = I_{n_1}$.

The encoding and decoding processes are:

I. Encoding:
$$m = sG_s G' \oplus r. \tag{1}$$

II. Decoding:
$$mP^{-1} = (sG_s SGP)P^{-1} \oplus rP^{-1} = (sG_s S)G \oplus r', \tag{2}$$

where $r'$ satisfies $W(r') = W(r) \leq t$, by means of fast decoding algorithm of Goppa code we can get $sG_sS$, then we get $x = sG_s$. Because $G_s$ is in standard form, the first k-bits of $x$ is the source message s. Suppose the parity check matrix of the linear code generated by $G_s = [I_k|A]$ is $H_s$, then $H_s = [-A^T|I_{n-k}]$. If the message Bob received has been changed in the channel, Bob will find that $xH_s^T \neq 0$.

---

[*] E-mail: yangli@gscas.ac.cn

Now let us consider the authentication scheme of quantum messages. A quantum message is a sequence of pure states:

$$M_k^{(n)} = \{\sum_m \alpha_m^{(i)}|m\rangle | i = 1,2,\cdots,n\}, \tag{3}$$

where $m = (m_1, m_2, \cdots, m_k) \in F_2^{(k)}$. Without loss of generality, we restrict our attention to the authentication of a pure state. Suppose Alice want to transmit a pure state $\sum_m \alpha_m |m\rangle$ to Bob with authentication, the encoding and decoding processes are:

I. Encoding:
   1. Compute:

$$\sum_m \alpha_m |m\rangle_I |0\rangle_{II} \to \sum_m \alpha_m |m\rangle_I |mG_sG'\rangle_{II} \to |0\rangle_I \sum_m \alpha_m |mG_sG'\rangle_{II}$$
$$\to |0\rangle_I \sum_m \alpha_m |mG_sG' \oplus e^{(i)}\rangle_{II} \tag{4}$$

   2. Send:

$$\sum_m \alpha_m |mG_sG' \oplus e^{(i)}\rangle_{II}. \tag{5}$$

II. Decoding:
   1. Compute:

$$\sum_m \alpha_m |mG_sG' \oplus e^{(i)}\rangle_{II} |0\rangle_{III} \to \sum_m \alpha_m |mG_sSG \oplus e^{(j)}\rangle_{II} |0\rangle_{III}$$
$$\to \sum_m \alpha_m |mG_sSG \oplus e^{(j)}\rangle_{II} |e^{(j)}H\rangle_{III} \tag{6}$$
$$\to \sum_m \alpha_m |mG_sSG \oplus e^{(j)}\rangle_{II} |s^{(j)}\rangle_{III}$$

   where $H$ is the parity check matrix of the Goppa code $C$.

   2. Measure the register *III* to get syndrome $s^{(j)}$, then to find $e^{(j)}$ via the fast decoding algorithm of the Goppa code $C$.
   3. Compute:

$$\begin{aligned}
&\hat{U}_{e^{(j)}} \sum_m \alpha_m \left| mG_s SG \oplus e^{(j)} \right\rangle_{II} |0\rangle_{III} |0\rangle_{IV} |0\rangle_V \\
&= \sum_m \alpha_m |mG_s SG\rangle_{II} |0\rangle_{III} |0\rangle_{IV} |0\rangle_V \\
&\to \sum_m \alpha_m |mG_s SG\rangle_{II} |0\rangle_{III} |mG_s S\rangle_{IV} |0\rangle_V \to |0\rangle_{II} |0\rangle_{III} \sum_m \alpha_m |mG_s S\rangle_{IV} |0\rangle_V \\
&\to |0\rangle_{II} |0\rangle_{III} \sum_m \alpha_m |mG_s\rangle_{IV} |0\rangle_V \to |0\rangle_{II} |0\rangle_{III} \sum_m \alpha_m |mG_s\rangle_{IV} \left| mG_s G_s^{-} \right\rangle_V \\
&\to |0\rangle_{II} |0\rangle_{III} \sum_m \alpha_m |mG_s\rangle_{IV} |m\rangle_V \to |0\rangle_{II} \sum_m \alpha_m |mG_s H_s\rangle_{III} |mG_s\rangle_{VI} |m\rangle_V \\
&= |0\rangle_{II} |0\rangle_{III} \sum_m \alpha_m |mG_s\rangle_{IV} |m\rangle_V \to |0\rangle_{II} |0\rangle_{III} |0\rangle_{IV} \sum_m \alpha_m |m\rangle_V
\end{aligned} \quad (7)$$

    4. Measure the third register to check whether it is in the state $|0\rangle$, then he get the message coming from Alice with authentication in the fifth register.

(There are some brief explanations of the computations between different quantum registers involved here in arXiv: quant-ph/0310076.)

If the matrix $G_s$ is public, the scheme is a public-key data integrity scheme. Data integrity is the property whereby data has not been altered in an unauthorized manner since the time it was created, transmitted, or stored by an authorized source [3]. The security of this public-key scheme is computational security. If the matrix $G_s$ is secret, the scheme is a hybrid data origin authentication (or message authentication) scheme.